\documentclass{article}
\usepackage[T1]{fontenc} 
\usepackage[utf8]{inputenc} 
\usepackage{ismir,amsmath,cite,url}
\usepackage{color}
\usepackage{booktabs}
\usepackage{tabularx}
\usepackage{boldline}
\usepackage{enumitem}
\usepackage[compact]{titlesec}
\usepackage{graphicx, subcaption}
\usepackage{caption}
\usepackage{amsfonts}

\definecolor{blue-violet}{rgb}{0.54, 0.17, 0.89}
\definecolor{ballblue}{rgb}{0.13, 0.67, 0.8}
\newcommand{\ELBO}{\text{ELBO}}
\newcommand{\thickhline}{\hlineB{2}}

\newcommand{\tianyaodeletes}[1]{}
\newcommand{\derek}[1]{{\color{cyan}Derek: #1}}

\newcommand{\indep}{\rotatebox[origin=c]{90}{$\models$}}

\title{Deep Music Analogy Via Latent Representation Disentanglement}

\multauthor
{Ruihan Yang$^1$ \hspace{.3cm} Dingsu Wang$^1$ \hspace{.3cm} Ziyu Wang$^1$ \hspace{.3cm} Tianyao Chen$^1$ \hspace{.3cm} Junyan Jiang$^{1,2}$ \hspace{.3cm} Gus Xia$^1$} {\\
$^1$ Music X Lab, NYU Shanghai \hspace{.3cm}
$^2$ Machine Learning Department, Carnegie Mellon University\\
{\tt\small$^1$\{ry649,dw1920,zz2417,tc2709,gxia\}@nyu.edu, $^2$junyanj@cs.cmu.edu}
}
\def\authorname{Ruihan Yang, et al}

\begin{document}
\maketitle
\begin{abstract}
\textit{Analogy-making} is a key method for computer algorithms to generate both natural and creative music pieces. In general, an analogy is made by partially transferring the music abstractions, i.e., high-level representations and their relationships, from one piece to another; however, this procedure requires \textit{disentangling} music representations, which usually takes little effort for musicians but is non-trivial for computers. Three sub-problems arise: extracting latent representations from the observation, disentangling the representations so that each part has a unique semantic interpretation, and mapping the latent representations back to actual music. In this paper, we contribute an explicitly-constrained conditional variational autoencoder (EC$^2$-VAE) as a unified solution to all three sub-problems. We focus on disentangling the \textit{pitch} and \textit{rhythm} representations of 8-beat music clips conditioned on chords. In producing music analogies, this model helps us to realize the imaginary situation of ``\textit{what if}'' a piece is composed using a different pitch contour, rhythm pattern, or chord progression by borrowing the representations from other pieces. Finally, we validate the proposed disentanglement method using objective measurements and evaluate the analogy examples by a subjective study.
\end{abstract}
\section{Introduction}\label{sec:introduction}

For intelligent systems, an effective way to generate high-quality art is to produce analogous versions of existing examples~\cite{hertzmann2001image}. In general, two systems are analogous if they share common abstractions, i.e., high-level representations and their relationships, which can be revealed by the paired tuples \textit{A : B :: C : D} (often spoken as A is to B as C is to D). For example, the analogy “the hydrogen atom is like our solar system” can be formatted as \textit{Nucleus : Hydrogen atom :: Sun : Solar system}, in which the shared abstraction is “a bigger part is the center of the whole system.” For generative algorithms, a clever shortcut is to make analogies by solving the problem of  “\textit{A : B :: C : ?}”. In the context of music generation, if A is the rhythm pattern of a very lyrical piece B, this analogy can help us realize the imaginary situation of “what if B is composed with a rather rapid and syncopated rhythm C” by preserving the pitch contours and the intrinsic relationship between pitch and rhythm. In the same fashion, other types of “what if” compositions can be created by simply substituting A and C with different aspects of music (e.g., chords, melody, etc.).

A great advantage of \textit{generation via analogy} is the ability to produce both \textit{natural} and \textit{creative} results. Naturalness is achieved by reusing the representations (high-level concepts such as ``image style'' and ``music pitch contour'') of human-made examples and the intrinsic relationship between the concepts, while creativity is achieved by recombining the representations in a novel way.  However, making meaningful analogies also requires \textit{disentangling} the representations, which is effortless for humans but non-trivial for computers. We already see that making analogies is essentially transferring the abstractions, not the observations --- simply copying the notes or samples from one piece to another would only produce a casual re-mix, not an analogous composition~\cite{gao2017towards}.

In this paper, we contribute an explicitly-constrained conditional variational autoencoder (EC$^2$-VAE), a conditional VAE with explicit semantic constraints on intermediate outputs of the network, as an effective tool for learning disentanglement. To be specific, the encoder extracts latent representations from the observations; the semantic constraints disentangle the representations so that each part has a unique interpretation, and the decoder maps the disentangled representations back to actual music while preserving the intrinsic relationship between the representations. In producing analogies, we focus on disentangling and transferring the \textit{pitch} and \textit{rhythm} representations of 8-beat music clips when chords are given as the condition (an extra input) of the model. We show that EC$^2$-VAE has three desired properties as a generative model. First, the disentanglement is \textit{explicitly coded}, i.e., we can specify which latent dimensions denote which semantic factors in the model structure. Second, the disentanglement does not sacrifice much of the reconstruction. Third, the learning does not require any analogous examples in the training phase, but the model is capable of making analogies in the inference phase. For evaluation, we propose a new metric and conduct a survey. Both objective and subjective evaluations show that our model significantly outperforms the baselines.  

\section{Related Work}
\subsection{Generation Via Analogy}
The history of generation via analogy can trace back to the studies of non-parametric “image analogies”~\cite{hertzmann2001image} and “playing Mozart by analogy” using case-based reasoning~\cite{widmer2003playing}. With recent breakthroughs in artificial neural networks, we see a leap in the quality of produced analogous examples using deep generative models, including music and image style transfer~\cite{dai2018music,gatys2015neural}, image-to-image translation~\cite{isola2017image}, attribute arithmetic~\cite{carter2017using}, and voice impersonation~\cite{gao2018voice}.

Here, we distinguish between two types of analogy algorithms. In a \textit{broad} sense, an analogy algorithm is any computational method capable of producing analogous versions of existing examples. A common and relatively easy approach is supervised learning, i.e., to directly learn the mapping between analogous items from labeled examples~\cite{isola2017image, tralie2018cover}. This approach requires little representation learning but needs a lot of labeling effort. Moreover, supervised analogy does not generalize well. For example, if the training analogous examples are all between lyrical melodies (the source domain) and syncopated melodies (the target domain), it would be difficult to create other rhythmic patterns, much less the manipulation of pitch contours. (Though improvements~\cite{zhu2017unpaired, kim2017learning, bouchacourt2018multi} have been made, weak supervision is still needed to specify the source and target domains.) On the other hand, a \textit{strict} analogy algorithm requires not only learning the representations but also disentangling them, which would allow the model to make domain-free analogies via the manipulation of any disentangled representations. Our approach belongs to this type.

\subsection{Representation Learning and Disentanglement}

Variational auto-encoders (VAEs)~\cite{kingma2013auto} and generative adversarial networks (GANs)~\cite{goodfellow2014generative} are so far the two most popular frameworks for music representation learning. Both use encoders (or discriminators) and decoders (or generators) to build a bi-directional mapping between the distributions of observation $x$ and latent representation $z$, and both generate new data via sampling from $p(z)$. For music representations, VAEs~\cite{roberts2018hierarchical, brunner2018midi, esling2018bridging, rh} have been a more successful tool so far compared with GANs~\cite{yu2017seqgan}, and our model is based on the previous study ~\cite{rh}.

The motivation of representation disentanglement is to better interpret the latent space generated by VAE, connecting certain parts of $z$ to semantic factors (e.g., age for face images, or rhythm for melody), which would enable a more controllable and interactive generation process.  InfoGAN~\cite{chen2016infogan} disentangles $z$ by encouraging the mutual information between $x$ and a subset of $z$. $\beta$-VAE~\cite{higgins2017beta} and its follow-up studies ~\cite{chen2018isolating, kim2018disentangling, rh} imposed various extra constraints and properties on $p(z)$. However, the disentanglement above are still \textit{implicit}, i.e., though the model separates the latent space into subparts, we cannot define their meanings beforehand and have to “check it out” via \textit{latent space traversal}~\cite{carter2017using}. In contrast, the disentanglement in Style-based GAN~\cite{karras2018style}, Disentangled Sequential Autoencoder~\cite{li2018disentangled}, and our EC$^2$-VAE are \textit{explicit}, i.e., the meanings of different parts of $z$ are defined by the model structure, so that the controlled generation is more precise and straightforward. The study Disentangled Sequential Autoencoder~\cite{li2018disentangled} is most related to our work and also deals with sequential inputs. Using a partially time-invariant encoder, it can approximately disentangle dynamic and static representations. Our model does not directly constrain $z$ but applies a loss to intermediate outputs associated with latent factors. Such an indirect but explicit constraint enables the model to further disentangle the representation into pitch, rhythm, and any semantic factors whose observation loss can be defined. As far as we know, this is the first disentanglement learning method tailored for music composition.

\section{Methodology}\label{sec:page_size}

In this section, we introduce the data representation and model design in detail. We focus on disentangling the latent representations of pitch and rhythm, the two fundamental aspects of composition, over the duration of 8-beat melodies. All data come from the Nottingham dataset~\cite{jukedeck2017}, regarding a $\frac{1}{4}$ beat as the shortest unit.

\subsection{Data Representation}\label{data}

Each 8-beat melody is represented as a sequence of 32 one-hot vectors each with 130 dimensions, where each vector denotes a $\frac{1}{4}$-beat unit. As in ~\cite{roberts2018hierarchical}, the first 128 dimensions denote the \textit{onsets} of MIDI pitches ranging from 0 to 127 with one unit of duration. The $129^{\text{th}}$ dimension is the \textit{holding} state for longer note duration, and the last dimension denotes \textit{rest}. We also designed a rhythm feature to constrain the intermediate output of the network. Each 8-beat rhythm pattern is also represented as a sequence of 32 one-hot vectors. Each vector has 3 dimensions, denoting: an onset of any pitch, a holding state, and rest. 

Besides, chords are given as a condition, i.e., an extra input, of the model. The chord condition of each 8-beat melody is represented as a chromagram with equal length, i.e., 32 multi-hot vectors each with 12 dimensions, where  each dimension indicates whether a pitch class is activated. 

\subsection{Model Architecture}

\begin{figure*}[htbp]
\centering
    \begin{subfigure}{0.45\textwidth}
        \centering
    	\includegraphics[height=9cm]{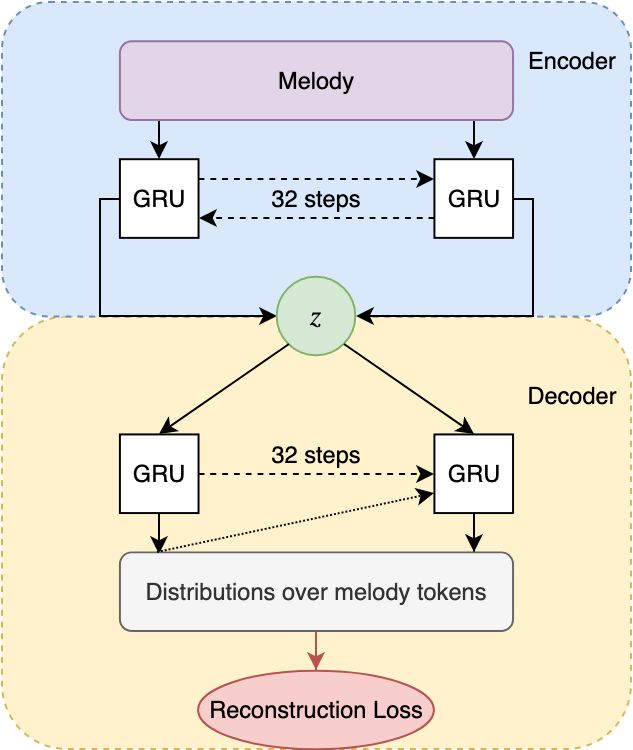}
    	\caption{Vanilla sequence VAE.}
		\label{fig:base}
    \end{subfigure}\hfill
    \begin{subfigure}{0.45\textwidth}
        \centering
    	\includegraphics[height=9cm]{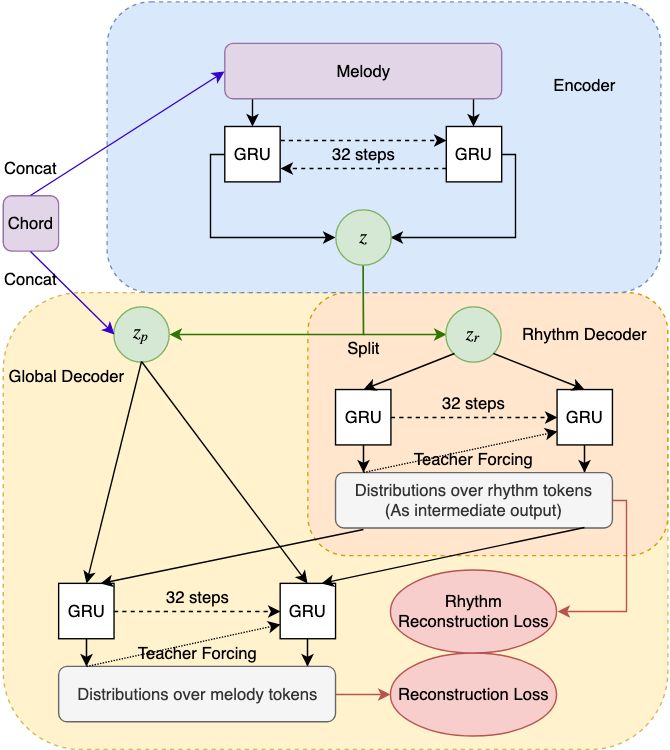}
    	\caption{EC$^2$-VAE model.}
		\label{fig:ecvae}
    \end{subfigure}
    \caption{A comparison between vanilla sequence VAE~\cite{rh} and our model with condition and disentanglement.}
    \label{fig:two_model}
\end{figure*}

Our model design is based on the previous studies of~\cite{roberts2018hierarchical, rh}, both of which used VAEs to learn the representations of fixed-length melodies. Figure \ref{fig:two_model} shows a comparison between the model architectures, where Figure \ref{fig:two_model}(\subref{fig:base}) shows the model designed in~\cite{rh} and Figure \ref{fig:two_model}(\subref{fig:ecvae}) shows the model design in this study. We see that both use bi-directional GRUs~\cite{chung2014empirical} (or LSTMs~\cite{hochreiter1997long}) as the encoders (in blue) to map each melody observation to a latent representation $z$, and both use uni-directional GRUs (or LSTMs) (with teacher forcing~\cite{toomarian1992learning} in the training phrase) as the decoders (in yellow) to reconstruct melodies from $z$.

The key innovation of our model design is to assign a part of the decoder (in orange) with a specific subtask: to disentangle the latent rhythm representation $z_r$ from the overall $z$ by explicitly encouraging the intermediate output of $z_r$ to match the rhythm feature of the melody. The other part of $z$ is therefore everything but rhythm and interpreted as the latent pitch representation, $z_p$. Note that this explicitly coded disentanglement technique is quite flexible --- \textit{we can use multiple subparts of the decoder to disentangle multiple semantically interpretable factors of $z$ simultaneously} as long as the intermediate outputs of the corresponding latent factors can be defined, and the model shown in Figure \ref{fig:two_model}(\subref{fig:ecvae}) is the simplest case of this family.

It is also worth noting that the new model uses chords as a condition for both the encoder and decoder. The advantage of chord conditioning is to free $z$ from storing chord-related information. In other words, the pitch information in $z$ is “detrended” by the underlying chord for better encoding and reconstruction. The cost of this design is that we cannot learn a latent distribution of chord progressions.

\subsubsection{Encoder}

A single layer bi-directional GRU with 32 time steps is used to model $Q_{\theta}(z|x,c)$, where $x$ is the melody input, $c$ is the chord condition, and $z$ is the latent representation. Chord conditions are given by concatenating with the input at each time step. 

\subsubsection{Decoder}

The global decoder models $P_\phi(x|z,c)$ by multiple layers of GRUs, each with 32 steps. For disentanglement, $z$ is splitted into two halves $z_p$ and $z_r$($z=\text{concat}[z_r , z_p]$), each being a 128-dimensional vector. As a subpart of the global decoder, the rhythm decoder models $P_{\phi_r}(r(x)|z)$ by a single layer GRU, where $r(x)$ is the rhythm feature of the melody. Meanwhile, the rhythm is concatenated with $z_p$ and chord condition as the input of the rest of the global decoder to reconstruct the melody. We used cross-entropy loss for both rhythm and melody reconstruction. Note that the overall decoder is supposed to learn non-trivial relationships between pitch and rhythm, rather than naively cutting a pitch contour by a rhythm pattern. 

\subsection{Theoretical Justification of the ELBO Objective with Disentanglement}

One concern about representation disentanglement techniques is that they sometimes sacrifice reconstruction power~\cite{kim2018disentangling}. In this section, we prove that our model does not suffer much of the disentanglement-reconstruction paradox, and the likelihood bound of our model is close to that of the original conditional VAE, and in some cases, equal to it.

Recall the Evidence Lower Bound (ELBO) objective function used by a typical conditional VAE~\cite{doersch2016tutorial} constraint on input sample $x$ with condition $c$:
\begin{align*}
    \ELBO(\phi, \theta) = & \ \mathbb{E}_Q[\log P_\phi(x|z,c)] \\
    &- \mathbb{KL}[Q_\theta(z|x,c)||P_\phi(z|c)] \leq \log P_\phi(x|c)
\end{align*}
For simplicity, $\mathcal{D}$ denotes $\mathbb{KL}[Q_\theta(z|x,c)||P_\phi(z|c)]$ in the rest of this section. If we see the intermediate rhythm output in Figure \ref{fig:two_model}(\subref{fig:ecvae}) as hidden variables of the whole network, the new ELBO objective of our model only adds the rhythm reconstruction loss based on the original one, resulting in a lower bound of the original ELBO. Formally,
\begin{align*}
    & \ \text{ELBO}^{\text{new}}(\phi, \theta) \\ 
    = & \ \mathbb{E}_Q[\log P_\phi(x|z,c)] - \mathcal{D} + \mathbb{E}_Q[\log P_{\phi_r}(r(x)|{z_r})]\\
    = & \ \text{ELBO}(\phi, \theta) + \mathbb{E}_Q[\log P_{\phi_r}(r(x)|{z_r})]
\end{align*}

\noindent where $\phi_r$ denotes parameters of the rhythm decoder. Clearly, ELBO$^\text{new}$ is a lower bound of the original ELBO because $\mathbb{E}_Q[\log P_{\phi_r}(r(x)|{z_r})] \leq 0$.

Moreover, if the rest of global decoder takes the original rhythm rather than the intermediate output of rhythm decoder as the input, the objective can be rewritten as:
\begin{align*}
    & \ \text{ELBO}^{\text{new}}(\phi, \theta)\\
    = & \ \mathbb{E}_Q[\underbrace{\log P_\phi(x|r(x), z_p, c) + \log P_\phi(r(x)|{z_r}, c)}_{\text{with } x \indep z_r |r(x),c \text{ and } r(x) \indep z_p | z_r, c}]- \mathcal{D} \\
    = & \ \mathbb{E}_Q[\log P_\phi(x, r(x) | z, c)] - \mathcal{D}
    \\
    = & \ \mathbb{E}_Q[\log P_\phi(x|z, c) + \log P_\phi(r(x)|x, z, c)] - \mathcal{D} \\
    = & \ \text{ELBO}(\phi, \theta)
\end{align*}

\noindent The second equal sign holds for a perfect disentanglement, and the last equal sign holds since $r(x)$ is decided by $x$, i.e., $ P_\phi(r(x)|x, z, c)=1$. In other words, we show that under certain assumptions $\text{ELBO}^{\text{new}}$ with disentanglement is identical to the ELBO.

\section{Experiments}
We present the objective metrics to evaluate the disentanglement in Section \ref{measure}, show several representative examples of generation via analogy in Section \ref{analogy}, and use subjective evaluations to rate the artistic aspects of the generated music in Section \ref{subject}. 

\subsection{Objective Measurements}\label{measure}

Upon a successful pitch-rhythm disentanglement, any changes in pitch of the original melody should not affect the latent rhythm representation much, and vice versa. Following this assumption, we developed two measurements to evaluate the disentanglement: 1) $\Delta z$ after transposition, which is more qualitative, and 2) F-score of an augmentation-based query, which is more quantitative. 

\subsubsection{Visualizing $\Delta z$ after transposition}

\begin{figure}[htbp]
    \centering
    \captionsetup{justification=centering}
	\includegraphics[width=\linewidth]{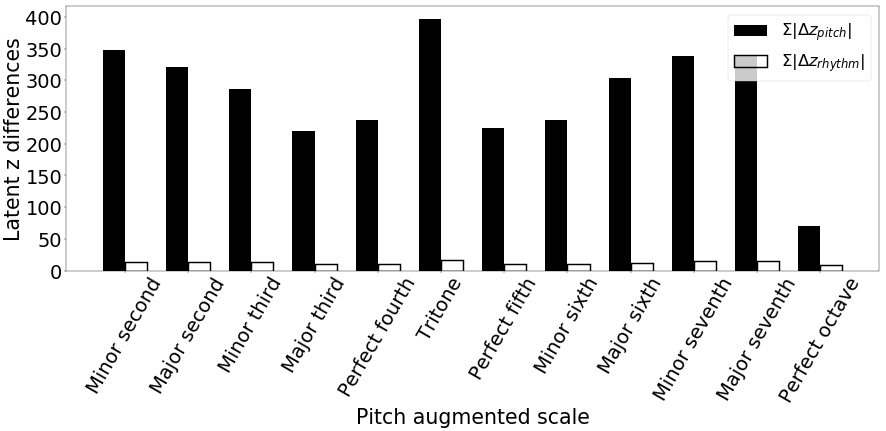}
	\caption{A comparison between $\Delta z_p$ and $\Delta z_r$ after transposition.}
	\label{fig:zplot}
\end{figure}

We define $F_i$ as the operation of transposing all the notes by $i$ semitones, and use the $L_1$-norm to measure the change in $z$. Figure \ref{fig:zplot} shows a comparison between $\Sigma |\Delta z_p|$ and $\Sigma |\Delta z_r|$ when we apply  $F_i$  to  a randomly chosen piece (where $i \in [1,12]$) while keeping the rhythm and underlying chord unchanged.

Here, the black bars stand for $\Sigma |\Delta z_p|$ and the white bars stand for the $\Sigma |\Delta z_r|$. It is conspicuous that when augmenting pitch, the change of $z_p$ is much larger than the change of $z_r$, which well demonstrates the success of the disentanglement.

It is also worth noting that the change of $z_p$ to a certain extent \textit{reflects human pitch perception}. Given a chord, the change in $z_p$ can be understood as the ``burden'' (or difficulty) to memorize (or encode) a transposed melody. We see that such burden is large for tritone (very dissonant), relatively small for major third, perfect fourth \& fifth (consonant), and very small for perfect octave.  

Due to the space limit, we only show the visualization of the latent space when changing the pitch. According to the data representation in Section \ref{data}, changing the rhythm feature of a melody would inevitably affect the pitch contour, which would lead to complex behavior of the latent space hard to interpret visually. We leave the discussion for future work but will pay more attention to the effect of the rhythm factor in Section \ref{subject}.

\subsubsection{$F$-score of Augmentation-based Query}

The explicitly coded disentanglement enables a new evaluation method from an \textit{information-retrieval} perspective. We regard the pitch-rhythm split in $z$ defined by the model structure as the \textit{reference} (the ground truth), the operation of factor-wise data augmentation (keeping the rhythm and only changing pitch randomly, or vice versa) as a \textit{query} in the latent space, and the actual latent dimensions having the largest variance caused by augmentation as the \textit{result set}. In this way, we can quantitatively evaluate our model in terms of precision, recall, and F-score.




\begin{figure}[htbp]
    \centering
    \captionsetup{justification=centering}
	\includegraphics[width=\linewidth]{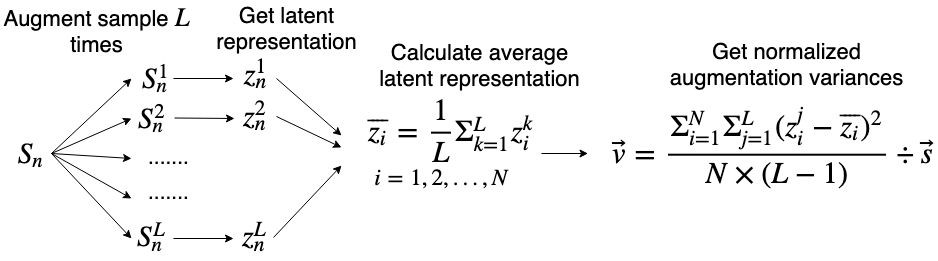}
	\caption{Evaluating the disentanglement by data augmentation.}
	\label{fig:quantitative-metric}
\end{figure}

\begin{table}[htbp]
\captionsetup{justification=centering}
\begin{tabularx}{\columnwidth}{l|rrr|rrr}
\thickhline
 & \multicolumn{3}{c|}{Pitch} & \multicolumn{3}{c}{Rhythm}  \\ \thickhline
 & pre.& rec. & $F$-s. & pre. & rec. & $F$-s.\\ \hline
EC$^2$-VAE & \textbf{0.88} & \textbf{0.88} & \textbf{0.88} & \textbf{0.80} & \textbf{0.80} & \textbf{0.80} \\
Random & 0.5 & 0.5 & 0.5 & 0.5 & 0.5 & 0.5 \\ \thickhline
\end{tabularx}
\caption {The evaluation results of pitch- and rhythm-wise augmentation-based query.}
\label{table:result}
\end{table}

Figure \ref{fig:quantitative-metric} shows the detailed query procedure, which is a modification of the evaluation method in~\cite{kim2018disentangling}. After pitch or rhythm augmentation for each sample, $\vec v$ is calculated as the average (across the samples) variance (across augmented versions) of the latent representations, normalized by the total sample variance $\vec s$. Then, we choose the first half (128 dimensions) with the largest variances as the result set. This precision, recall and F-score of this augmentation-based query result is shown in Table \ref{table:result}. (Here, precision and recall are identical since the size of the result set equals the dimensionality of $z_p$ and $z_r$.) As this is the first tailored metric for explicitly coded disentanglement, we use random guess as our baseline.

\subsection{Examples of Generation via Analogy}\label{analogy}

\begin{figure}[!h]
    \centering
	\includegraphics[width=\linewidth]{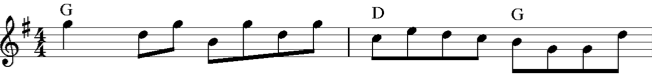}
	\caption{The source melody.}
	\label{fig:src}
\end{figure}

We present several representative “what if” examples by swapping or interpolating the latent representations of different pieces. Throughout this section, we use the following example (shown in Figure \ref{fig:src}), an 8-beat melody from the Nottingham Dataset~\cite{jukedeck2017} as the source, and the target rhythm or pitch will be borrowed from other pieces. (MIDI demos are available at \url{https://github.com/cdyrhjohn/Deep-Music-Analogy-Demos}.)

\subsubsection{Analogy by replacing $z_p$}

Two examples are presented. In both cases, the latent pitch representation and the chord condition of the source melody are replaced with new ones from other pieces. In other words, the model answers the analogy question: \textit{“source’s pitch : source melody :: target’s pitch : ?”}

\begin{figure}[htbp]
    \begin{subfigure}{\columnwidth}
    	\includegraphics[width=\columnwidth]{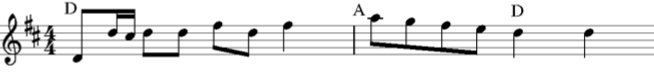}
    	\caption{Target’s pitch and chord.}
    	\label{fig:pt1}
    \end{subfigure}
    \par
    \begin{subfigure}{\columnwidth}
        \captionsetup{justification=centering}
    	\includegraphics[width=\columnwidth]{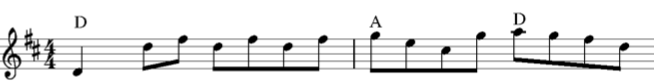}
    	\caption{The generated target music, using the pitch and chord from (a) and the rhythm from the source.}
    	\label{fig:pg1}
    \end{subfigure}
    \caption{The $1^\text{st}$ example of analogy via replacing $z_p$.}
    \label{fig:p1}
\end{figure}

Figure \ref{fig:p1} shows the first example, where Figure \ref{fig:p1}(\subref{fig:pt1}) shows the piece from which the pitch and chords are borrowed, and Figure \ref{fig:p1}(\subref{fig:pg1}) shows the generated melody. From Figure \ref{fig:p1}(\subref{fig:pt1}), we see the target melody is in a different key (D major) with a larger pitch range than the source and a big pitch jump in the beginning. From Figure \ref{fig:p1}(\subref{fig:pg1}), we see the generated new melody captures such pitch features while keeping the rhythm of the source unchanged.

\begin{figure}[htbp]
    \begin{subfigure}{\columnwidth}
        \captionsetup{justification=centering}
    	\includegraphics[width=\columnwidth]{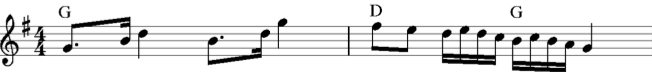}
    	\caption{Target’s pitch and chord.}
    	\label{fig:pt2}
    \end{subfigure}
    \par
    \begin{subfigure}{\columnwidth}
        \captionsetup{justification=centering}
    	\includegraphics[width=\columnwidth]{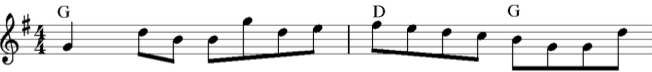}
    	\caption{The generated target, using the pitch and chord from (a) and the rhythm from the source.}
    	\label{fig:pg2}
    \end{subfigure}
    \caption{The $2^\text{nd}$ analogy example via replacing $z_p$.}
    \label{fig:p2}
\end{figure}

Figure \ref{fig:p2} shows another example, whose subplots share the same meanings with the previous one. From Figure \ref{fig:p2}(\subref{fig:pt2}), we see the first measure of the target’s melody is a broken chord of Gmaj, while the second measure is the G major scale. From Figure \ref{fig:p2}(\subref{fig:pg2}), we see the generated new melody captures these pitch features. Moreover, it retains the source’s rhythm and ignores the dotted eighth and sixteenth notes in Figure \ref{fig:p2}(\subref{fig:pt2}).

\subsubsection{Analogy by replacing $z_r$}

\begin{figure}[htbp]
    \begin{subfigure}{\columnwidth}
    	\includegraphics[width=\columnwidth]{figs/p_target_1.png}
    	\caption{Target’s rhythm pattern.}
    	\label{fig:rt1}
    \end{subfigure}
    \par
    \begin{subfigure}{\columnwidth}
        \captionsetup{justification=centering}
    	\includegraphics[width=\columnwidth]{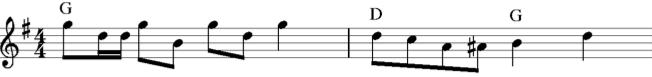}
    	\caption{The generated target music, using the rhythm of (a) while keeping source’s pitch and chord.}
    	\label{fig:rg1}
    \end{subfigure}
    \caption{The $1^\text{st}$ example of analogy via replacing $z_r$.}
    \label{fig:r1}
\end{figure}

Similar to the previous section, this section shows two example answers to the question: “\textit{source’s rhythm : source melody :: target’s rhythm : ?}” by replacing $z_r$.
Figure \ref{fig:r1} shows the first example, where Figure \ref{fig:r1}(\subref{fig:rt1}) contains the new rhythm pattern quite different from the source, and Figure \ref{fig:r1}(\subref{fig:rg1}) is the generated target. We see that Figure \ref{fig:r1}(\subref{fig:rg1}) perfectly inherited the new rhythm pattern and made  minor but novel modifications based on the source’s pitch.

\begin{figure}[htbp]
    \begin{subfigure}{\columnwidth}
    	\includegraphics[width=\columnwidth]{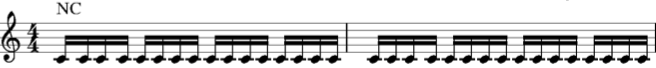}
    	\caption{Target’s rhythm pattern.}
    	\label{fig:rt2}
    \end{subfigure}
    \par
    \begin{subfigure}{\columnwidth}
        \captionsetup{justification=centering}
    	\includegraphics[width=\columnwidth]{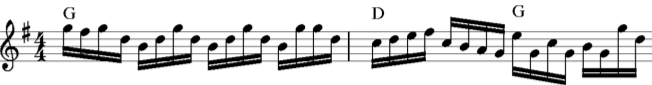}
    	\caption{The generated target music, using the rhythm of (a) while keeping source’s pitch and chord.}
    	\label{fig:rg2}
    \end{subfigure}
    \caption{The $2^\text{nd}$ analogy example via replacing $z_r$.}
    \label{fig:r2}
\end{figure}

Figure \ref{fig:r2} shows a more extreme case, in which Figure \ref{fig:r2}(\subref{fig:rt2}) contains only 16th notes of the same pitch. Again, we see the generated target in Figure \ref{fig:r2}(\subref{fig:rg2}) maintains the source’s pitch contour while matching the given rhythm pattern.

\subsubsection{Analogy by Replacing Chord}

\begin{figure}[!h]
    \begin{subfigure}{\columnwidth}
    	\includegraphics[width=\columnwidth]{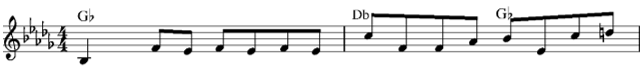}
    	\captionsetup{justification=centering}
    	\caption{Changing all the chords down a semitone, resulting in the key change from G major to Bb minor.}
    	\label{fig:cg1}
    \end{subfigure}
    \par
    \begin{subfigure}{\columnwidth}
    	\includegraphics[width=\columnwidth]{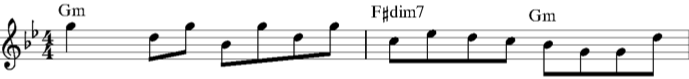}
    	\captionsetup{justification=centering}
    	\caption{Changing the key from G major to G minor.}
    	\label{fig:cg2}
    \end{subfigure}
    \caption{Two examples of replacing the original chord.}
    \label{fig:cg}
\end{figure}

Though chord is not our main focus, here we show two analogy examples in Figure \ref{fig:cg} to answer “what if” the source melody is composed using some other chord progressions. Figure \ref{fig:cg}(\subref{fig:cg1}) shows an example where the key is Bb minor. An interesting observation is the new melody contour indeed adds some reasonable modification (e.g. flipping the melody) rather than simply transposing down all the notes. It brings us a little sense of Jazz. Figure \ref{fig:cg}(\subref{fig:cg2}) shows an example where the key is changed from G major to G minor. We see melody also naturally transforms from major mode to minor mode.

\subsubsection{Two-way Pitch-Rhythm Interpolation}

\begin{figure}[htbp]
    \centering
    \captionsetup{justification=centering}
	\includegraphics[width=\columnwidth]{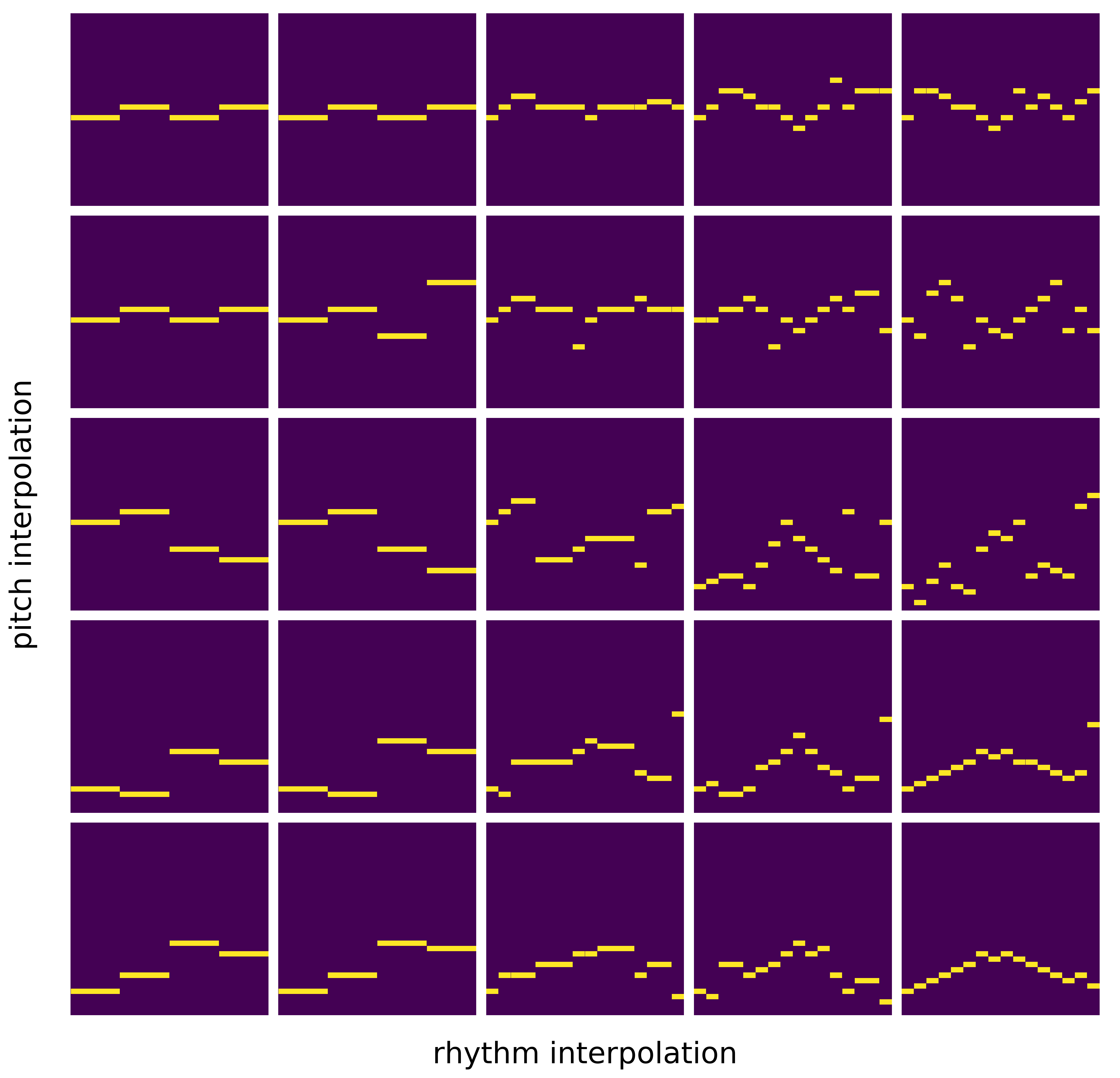}
	\caption{An illustration of two-way interpolation.}
	\label{fig:interp}
\end{figure}

The disentanglement also enables a smooth transition from one music to another. Figure \ref{fig:interp} shows an example of two-way interpolation, i.e., a traversal over a subspace of the learned latent representations $z_r$ and $z_p$ along 2 axes respectively, while keeping the chord as NC (no chord). Here, each square is a piano-roll of an 8-beat music. The top-left (source) and bottom-right (target) squares are two samples created manually and everything else is generated by interpolation using SLERP~\cite{watt1992advanced}. Note that the rhythmic changes are primarily observed moving along the ``rhythm interpolation'' axis, and likewise for pitch and the vertical ``pitch interpolation'' axis.

\subsection{Subjective Evaluation}\label{subject}
Besides objective measurement, we conducted a subjective survey to evaluate the quality of generation via analogy. We focus on changing the rhythm factors of existing music since this operation leads to an easier identification of the source melodies.

Each subject listened to two groups of five pieces each. All the pieces had the same length (64 beats at 120 bpm). Within each group, one piece was an original, human-composed piece from the Nottingham dataset, having a lyrical melody consisting of longer notes. The remaining four pieces were variations upon the original with more rapid rhythms consisting of 8$^\text{th}$ and 16$^\text{th}$ notes. Two of the variations were produced in a rule-based fashion by naively cutting the notes in the original into shorter subdivisions, serving as the \textit{baseline}. The other two variations were generated with our EC$^2$-VAE by merging the $z_p$ of the original piece and the $z_r$ decoded from two pieces having the same rhythm pattern as the baselines but with all notes replaced with ``do'' (similar to  Figure~\ref{fig:r2}(a)). The subjects always listened to the original first, and the order of the variations was randomized. In sum, we compare three versions of music: 1) the original piece, 2) the variation created by the baseline, and, 3) the variation created by our algorithm. The subjects were asked to rate each sample on a 5-point scale from 1 (very low) to 5 (very high) according to three criteria:

\begin{enumerate}[noitemsep]
    \item \textit{Creativity}: how creative the composition is.  
    \item \textit{Naturalness}: how human-like the composition is.
    \item \textit{Overall musicality}.
\end{enumerate}

\begin{figure}[htbp]
    \centering
	\includegraphics[width=\linewidth]{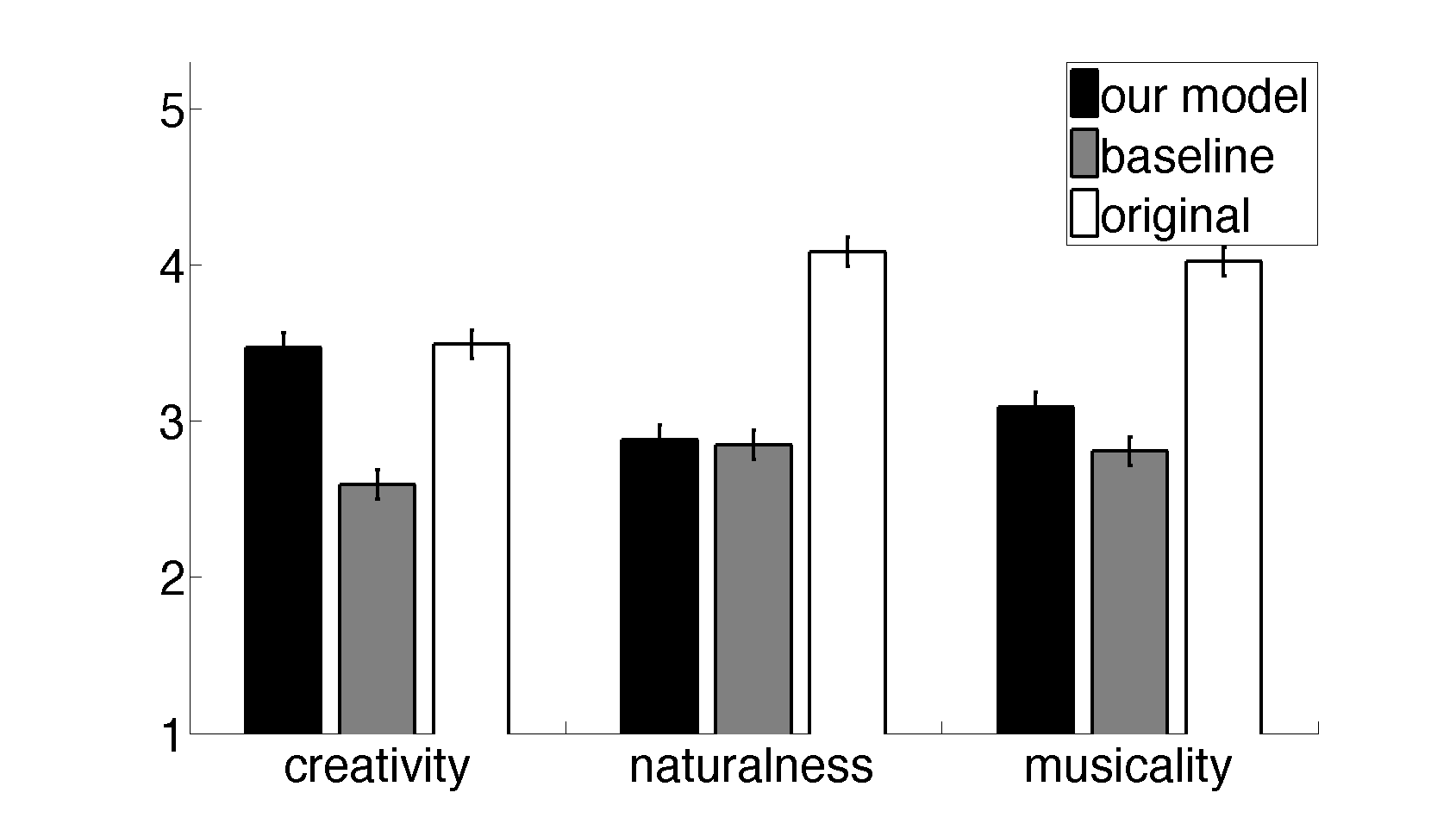}
	\caption{Subjective evaluation results.}
	\label{fig:se}
\end{figure}

A total of 30 subjects (16 female and 14 male) participated in the survey. Figure~\ref{fig:se} shows the results, where the heights of bars represent means of the ratings the and error bars represent the MSEs computed via within-subject ANOVA~\cite{scheffe1999analysis}. The result shows that our model performs significantly better than the rule-based baseline in terms of creativity and musicality ($p < 0.05$), and marginally better in terms of naturalness. Our proposed method is even comparable to the original music in terms of creativity, but remains behind human composition in terms of the other two criteria.

\section{Conclusion}
In conclusion, we contributed an explicitly-constrained conditional variational autoencoder (EC$^2$-VAE) as an effective disentanglement learning model. This model generates new music via making analogies, i.e., to answer the imaginary situation of “what if” a piece is composed using different pitch contours, rhythm patterns, and chord progressions via replacing or interpolating the disentangled representations. Experimental results showed that the disentanglement is successful and the model is able to generate interesting and musical analogous versions of existing music.  We see this study a significant step in music understanding and controlled music generation. The model also has the potential to be generalized to other domains, shedding light on the general scenario of generation via analogy.

\section{Acknowledgement}
We thank Yun Wang, Zijian Zhou and Roger Dannenberg for the in-depth discussion on music disentanglement and analogy. This work is partially supported by the Eastern Scholar Program of Shanghai.

\bibliography{ISMIR2019}

\begin{thebibliography}{10}

\bibitem{bouchacourt2018multi}
Diane Bouchacourt, Ryota Tomioka, and Sebastian Nowozin.
\newblock Multi-level variational autoencoder: Learning disentangled
  representations from grouped observations.
\newblock In {\em Thirty-Second AAAI Conference on Artificial Intelligence},
  2018.

\bibitem{brunner2018midi}
Gino Brunner, Andres Konrad, Yuyi Wang, and Roger Wattenhofer.
\newblock Midi-vae: Modeling dynamics and instrumentation of music with
  applications to style transfer.
\newblock {\em arXiv preprint arXiv:1809.07600}, 2018.

\bibitem{carter2017using}
Shan Carter and Michael Nielsen.
\newblock Using artificial intelligence to augment human intelligence.
\newblock {\em Distill}, 2(12):e9, 2017.

\bibitem{chen2018isolating}
Ricky~TQ Chen, Xuechen Li, Roger Grosse, and David Duvenaud.
\newblock Isolating sources of disentanglement in vaes.
\newblock NIPS, 2018.

\bibitem{chen2016infogan}
Xi~Chen, Yan Duan, Rein Houthooft, John Schulman, Ilya Sutskever, and Pieter
  Abbeel.
\newblock Infogan: Interpretable representation learning by information
  maximizing generative adversarial nets.
\newblock In {\em Advances in neural information processing systems}, pages
  2172--2180, 2016.

\bibitem{chung2014empirical}
Junyoung Chung, Caglar Gulcehre, KyungHyun Cho, and Yoshua Bengio.
\newblock Empirical evaluation of gated recurrent neural networks on sequence
  modeling.
\newblock {\em arXiv preprint arXiv:1412.3555}, 2014.

\bibitem{dai2018music}
Shuqi Dai, Zheng Zhang, and Gus~G Xia.
\newblock Music style transfer: A position paper.
\newblock {\em arXiv preprint arXiv:1803.06841}, 2018.

\bibitem{doersch2016tutorial}
Carl Doersch.
\newblock Tutorial on variational autoencoders.
\newblock {\em arXiv preprint arXiv:1606.05908}, 2016.

\bibitem{esling2018bridging}
Philippe Esling, Axel Chemla-Romeu-Santos, and Adrien Bitton.
\newblock Bridging audio analysis, perception and synthesis with
  perceptually-regularized variational timbre spaces.
\newblock In {\em Proceedings of the 19th International Society for Music
  Information Retrieval Conference, ISMIR}, pages 23--27, 2018.

\bibitem{jukedeck2017}
E.~Foxley.
\newblock Nottingham database, 2011.

\bibitem{gao2017towards}
Y~Gao.
\newblock {\em Towards neural music style transfer}.
\newblock PhD thesis, Master Thesis, New York University. https://github.
  com/821760408-sp/the~…, 2017.

\bibitem{gao2018voice}
Yang Gao, Rita Singh, and Bhiksha Raj.
\newblock Voice impersonation using generative adversarial networks.
\newblock In {\em 2018 IEEE International Conference on Acoustics, Speech and
  Signal Processing (ICASSP)}, pages 2506--2510. IEEE, 2018.

\bibitem{gatys2015neural}
Leon~A Gatys, Alexander~S Ecker, and Matthias Bethge.
\newblock A neural algorithm of artistic style.
\newblock {\em arXiv preprint arXiv:1508.06576}, 2015.

\bibitem{goodfellow2014generative}
Ian Goodfellow, Jean Pouget-Abadie, Mehdi Mirza, Bing Xu, David Warde-Farley,
  Sherjil Ozair, Aaron Courville, and Yoshua Bengio.
\newblock Generative adversarial nets.
\newblock In {\em Advances in neural information processing systems}, pages
  2672--2680, 2014.

\bibitem{hertzmann2001image}
Aaron Hertzmann, Charles~E Jacobs, Nuria Oliver, Brian Curless, and David~H
  Salesin.
\newblock Image analogies.
\newblock In {\em Proceedings of the 28th annual conference on Computer
  graphics and interactive techniques}, pages 327--340. ACM, 2001.

\bibitem{higgins2017beta}
Irina Higgins, Loic Matthey, Arka Pal, Christopher Burgess, Xavier Glorot,
  Matthew Botvinick, Shakir Mohamed, and Alexander Lerchner.
\newblock beta-vae: Learning basic visual concepts with a constrained
  variational framework.
\newblock In {\em International Conference on Learning Representations},
  volume~3, 2017.

\bibitem{hochreiter1997long}
Sepp Hochreiter and J{\"u}rgen Schmidhuber.
\newblock Long short-term memory.
\newblock {\em Neural computation}, 9(8):1735--1780, 1997.

\bibitem{isola2017image}
Phillip Isola, Jun-Yan Zhu, Tinghui Zhou, and Alexei~A Efros.
\newblock Image-to-image translation with conditional adversarial networks.
\newblock In {\em Proceedings of the IEEE conference on computer vision and
  pattern recognition}, pages 1125--1134, 2017.

\bibitem{karras2018style}
Tero Karras, Samuli Laine, and Timo Aila.
\newblock A style-based generator architecture for generative adversarial
  networks.
\newblock {\em arXiv preprint arXiv:1812.04948}, 2018.

\bibitem{kim2018disentangling}
Hyunjik Kim and Andriy Mnih.
\newblock Disentangling by factorising.
\newblock {\em arXiv preprint arXiv:1802.05983}, 2018.

\bibitem{kim2017learning}
Taeksoo Kim, Moonsu Cha, Hyunsoo Kim, Jung~Kwon Lee, and Jiwon Kim.
\newblock Learning to discover cross-domain relations with generative
  adversarial networks.
\newblock In {\em Proceedings of the 34th International Conference on Machine
  Learning-Volume 70}, pages 1857--1865. JMLR. org, 2017.

\bibitem{kingma2013auto}
Diederik~P Kingma and Max Welling.
\newblock Auto-encoding variational bayes.
\newblock {\em arXiv preprint arXiv:1312.6114}, 2013.

\bibitem{li2018disentangled}
Yingzhen Li and Stephan Mandt.
\newblock Disentangled sequential autoencoder.
\newblock {\em arXiv preprint arXiv:1803.02991}, 2018.

\bibitem{roberts2018hierarchical}
Adam Roberts, Jesse Engel, Colin Raffel, Curtis Hawthorne, and Douglas Eck.
\newblock A hierarchical latent vector model for learning long-term structure
  in music.
\newblock {\em arXiv preprint arXiv:1803.05428}, 2018.

\bibitem{scheffe1999analysis}
Henry Scheffe.
\newblock {\em The analysis of variance}, volume~72.
\newblock John Wiley \& Sons, 1999.

\bibitem{toomarian1992learning}
Nikzad~Benny Toomarian and Jacob Barhen.
\newblock Learning a trajectory using adjoint functions and teacher forcing.
\newblock {\em Neural Networks}, 5(3):473--484, 1992.

\bibitem{tralie2018cover}
Christopher~J Tralie.
\newblock Cover song synthesis by analogy.
\newblock {\em arXiv preprint arXiv:1806.06347}, 2018.

\bibitem{watt1992advanced}
Alan Watt and Mark Watt.
\newblock Advanced animation and bendering techniques.
\newblock 1992.

\bibitem{widmer2003playing}
Gerhard Widmer and Asmir Tobudic.
\newblock Playing mozart by analogy: Learning multi-level timing and dynamics
  strategies.
\newblock {\em Journal of New Music Research}, 32(3):259--268, 2003.

\bibitem{rh}
Ruihan Yang, Tianyao Chen, Yiyi Zhang, and Gus Xia.
\newblock Inspecting and interacting with meaningful music representations
  using vae.
\newblock {\em arXiv preprint arXiv:1904.08842}, 2019.

\bibitem{yu2017seqgan}
Lantao Yu, Weinan Zhang, Jun Wang, and Yong Yu.
\newblock Seqgan: Sequence generative adversarial nets with policy gradient.
\newblock In {\em Thirty-First AAAI Conference on Artificial Intelligence},
  2017.

\bibitem{zhu2017unpaired}
Jun-Yan Zhu, Taesung Park, Phillip Isola, and Alexei~A Efros.
\newblock Unpaired image-to-image translation using cycle-consistent
  adversarial networks.
\newblock In {\em Proceedings of the IEEE international conference on computer
  vision}, pages 2223--2232, 2017.

\end{thebibliography}

%
%
%
%

\end{document}